\documentclass[12pt,tightenlines,eqsecnum,floats,showpacs,nofootinbib,amsmath,amssymb,aps,prd]{revtex4}

\usepackage{setspace}
\usepackage{amsmath,amssymb,amsfonts,amsthm,latexsym}
\usepackage{graphicx}
\usepackage{enumerate} 
\usepackage{colordvi} 

\def\be{\begin{equation}}
\def\ee{\end{equation}}
\def\ba{\begin{eqnarray}}
\def\ea{\end{eqnarray}}
\def\f{\frac}

\def\h{\hat}

\def\L{\mathcal{L}}
\def\H{\mathcal{H}}
\def\Hp{\mathcal{H}_{\rm phy}}
\def\Hpo{\mathcal{H}_{\rm phy}^o}

\def\inter{{\rm int}}

\def\r{\rho}
\def\rp{\rho_{\rm Pl}}

\def\l{\lambda}
\def\lp{\ell_{\rm Pl}}
\def\T{T}
\def\pT{P_{(\T)}}
\def\dd{{\rm d}}
\def\v{\nu}
\def\vp{\varphi}
\def\pphi{\pi_{(\varphi)}}

\newcommand{\bra}{\langle}
\newcommand{\ket}{\rangle}

\newcommand{\Z}{{\mathbb{Z}}}
\newcommand{\R}{{\mathbb{R}}}

\newcommand{\kv}{{\vec{k}}}
\newcommand{\xv}{{\vec{x}}}
\newcommand{\q}{{\hat{q}}}
\newcommand{\p}{{\hat{p}}}

\begin{document}


\title{Quantum field theory on a cosmological, quantum space-time}
\author{Abhay Ashtekar${}^1$}
\email{ashtekar@gravity.psu.edu}
\author{Wojciech Kaminski${}^{1,2}$}
\email{Wojciech.Kaminski@fuw.edu.pl}
\author{Jerzy Lewandowski${}^{1,2}$}
\email{jerzy.lewandowski@fuw.edu.pl}
\affiliation{${}^1$Institute for Gravitation and the Cosmos \&
Physics Department, Penn State, University Park, PA 16802, U.S.A.\\
${}^{2}$ Instytut Fizyki Teoretycznej Uniwersytet Warszawski,
ul.Hoza 69, PL-00 681 Warsaw, Poland}

\begin{abstract}

In loop quantum cosmology,
Friedmann-LeMa\^{\i}tre-Robertson-Walker (FLRW) space-times arise
as well-defined approximations to specific \emph{quantum}
geometries. We initiate the development of a quantum theory of
test scalar fields on these quantum geometries. Emphasis is on the
new conceptual ingredients required in the transition from
classical space-time backgrounds to quantum space-times. These
include a `relational time' a la Leibniz, the emergence of the
Hamiltonian operator of the test field from the quantum constraint
equation, and ramifications of the quantum fluctuations of the
background geometry on the resulting dynamics. The familiar
quantum field theory on classical FLRW models arises as a
well-defined reduction of this more fundamental theory.

\end{abstract}

\maketitle
\section{Introduction}
\label{s1}

Quantum field theory (QFT) on classical
Friedmann-LeMa\^{\i}tre-Robertson-Walker (FLRW) space-times is well
developed and has had remarkable success in accounting for structure
formation in inflationary cosmologies (see, e.g., \cite{mw}). In
this analysis one assumes that the background space-time is
adequately described by classical general relativity. During the
inflationary era, this assumption is reasonable because, e.g., in
the standard scenarios the matter density $\r$ even at the onset of
inflation is less than $10^{-10}\, \rp$, where $\rp$ is the Planck
density. However, even in an eternal inflation, the underlying
classical space-time has a big bang singularity \cite{bgv}. The
theory is thus incomplete. In particular, the presence of this
singularity makes it awkward to introduce initial conditions, e.g.,
on the quantum state of matter.

To know what really happened in the Planck regime near the
singularity, we need a quantum theory of gravity. While a fully
satisfactory quantum gravity theory is still not available, over
the past 2-3 years, loop quantum cosmology (LQC) has provided a
number of concrete results on this Planck scale physics. (For recent
reviews, see, e.g., \cite{aa-badhonef,aa-saloniki}.) LQC is a
symmetry reduced version of loop quantum gravity (LQG)
\cite{alrev,crbook,ttbook}), a non-perturbative, background
independent approach to the unification of general relativity and
quantum physics. Here, space-time geometry is treated quantum
mechanically from the start. In the symmetry reduced cosmological
models these quantum geometry effects create a new repulsive force
when space-time curvature enters the Planck regime. The force is so
strong that the big bang is replaced by a specific type of quantum
bounce \cite{mb1,abl,aps1,aps2,aps3,acs,cs1,kl}. The force rises
very quickly once $\r$ exceeds
$\sim 0.01\,\rp$ to cause the bounce but also dies very quickly
after the bounce, once the density falls below this value.
Therefore, the quantum space-time of LQC is very well approximated
by the space-time continuum of general relativity once the curvature
falls below the Planck scale. This scenario is robust in the sense
that it is borne out for k=0 models with or without a cosmological
constant \cite{bp,ap}, k=1 closed models, and
\cite{apsv,warsaw}, k=-1 open models%
\footnote{The current treatment of the k=-1 models is not entirely
satisfactory because it regards the extrinsic curvature $K_a^i$ as
a connection and relies on holonomies constructed from it.
However, a closer examination shows that this is not necessary
\cite{awe4}.}
\cite{kv}. The Bianchi I model which incorporates anisotropies
\cite{awe2} and the k=0 model with an inflationary potential with
phenomenologically viable parameters \cite{aps4}.

In this paper, we will use the detailed quantum geometries that have
been constructed in LQC for the k=0, $\Lambda =0$, FLRW models with
a massless scalar field as a source. The full physical Hilbert space
of LQC is infinite dimensional. Every physical state undergoes a
quantum bounce in a precise sense \cite{acs}. However, for physical
applications of interest here, we will consider only those states
which are sharply peaked on a classical geometry at \emph{some} late
time and follow their evolution. Surprisingly, LQC predicts that
dynamics of these states is well approximated by certain `effective
trajectories' \cite{jw,vt} in the gravitational phase space at
\emph{all} times, including the bounce point \cite{aps3,acs}. As one
would expect, this effective trajectory departs sharply from the
solution to Einstein's equation near the bounce. However it does
define a smooth space-time metric, but its coefficients now involve
$\hbar$. These quantum corrections are extremely large in the Planck
regime but, as indicated above, die off quickly and the effective
space-time is indistinguishable from
the classical FLRW solution in the low curvature region.%
\footnote{The availability of a singularity free effective
space-time can be extremely useful. For example, it has enabled one
to show that, although Bousso's covariant entropy bound \cite{rb} is
violated very near the singularity in classical general relativity,
it is respected in the quantum space-time of LQC.}

Thus, LQC provides specific, well-defined quantum geometries from
which FLRW space-times emerge away from the Planck scale. At a
fundamental level, one does not have a single classical metric but
rather a probability amplitude for various metrics. So, the
question naturally arises: \emph{How do quantum fields propagate
on these quantum geometries?}

Availability of a satisfactory quantum theory of fields on a quantum
geometry would provide new perspectives in a number of directions.
First, it could provide a coherent theory of structure formation
from first principles.
For example, one may be able to specify the initial conditions
either in the infinite past where quantum space-time is well
approximated by a flat classical geometry, or, at the bounce point
which now replaces the big-bang. Second, the theory is also of
considerable importance from a more general conceptual
perspective. For, it should provide a bridge between quantum
gravity and QFT in curved space-times. What precisely are the
implications of the quantum fluctuations of geometry on the
dynamics of other quantum fields? What, in particular, are the
consequences of light cone fluctuations? Finally, this theory
would lead to a rich variety of new avenues in mathematical
physics. How is the relational time of quantum gravity related to
the more familiar choices of time one makes in QFT in curved
space-times? How do the standard anomalies of QFT on classical
background geometries `lift' up to QFT on quantum geometries? What
precisely are the approximations that enable one to pass from
quantum QFT on quantum geometries to those on classical
geometries?

The purpose of this paper is to provide the first steps to
addressing these important issues. More precisely, we will present
the basics of a framework to describe \emph{test} quantum fields on
the quantum FLRW geometries provided by LQC.

QFT in curved space-times has been developed in two directions. The
first is the more pragmatic approach that cosmologists have
developed to study structure formation, particle creation by given
gravitational fields, and their back reaction on the geometry (see,
e.g., \cite{mw}). Here, one uses the background geometry to make a
mode decomposition and regards the quantum field as an assembly of
oscillators. Typically, one focuses on one mode (or a finite number
of modes) at a time and ignores the difficult functional analytical
issues associated with the fact that the field in fact has an
infinite number of degrees of freedom. The second direction is the
more mathematical, algebraic approach that provides a conceptually
complete home for the subject (see, e.g., \cite{rmw,bsk}). Here the
focus is on the structure of operator algebras, constructed
`covariantly' using the background space-time geometry. States are
treated as suitably regular positive linear functionals on the
algebras. Not only is there no mode decomposition but one does not
tie oneself to any one Hilbert space. Our long range goal is to
generalize both sets of analyses to quantum space-times.

In this paper we will make a beginning by following the more
pragmatic approach: As in the literature on cosmology, we will use
mode decomposition. However in this analysis, our emphasis will be
on conceptual issues. First, in LQC one is led to a relational
dynamics because there is no background space-time. More precisely,
one `deparametrizes' the theory: the massless scalar field $\T$
---the matter source in the background space-time--- is treated as
the `evolution parameter' with respect to which the physical degrees
of freedom ---the density, volume, anisotropies and other matter
fields, if any--- evolve. Therefore, in QFT on FLRW quantum
geometries, it is natural to continue to use $\T$ as time. In QFT on
classical FLRW space-times, on the other hand, one generally uses
the conformal or proper time as the evolution parameter. We will
resolve this conceptual tension. Second, in the quantum gravity
perspective, dynamics is encoded in the quantum constraint equation.
In QFT on a classical FLRW geometry, on the other hand, dynamics of
the test quantum field is generated by a Hamiltonian. We will show
how this Hamiltonian naturally emerges from the quantum constraint
in a suitable approximation. The analysis is quite intricate because
it involves different notions of time (or, equivalently, lapse
fields) at different stages. Finally we will be able to pin-point
the implications of the quantum fluctuations of geometry on the
dynamics of the test quantum field. This discussion will, in turn,
enable us to spell out the approximations that are essential to pass
from the QFT on a quantum FLRW geometry to that on its classical
counterpart.

The paper is organized as follows. In section \ref{s2} we summarize
key properties of quantum \emph{space-time} geometries that emerge
from LQC and recall the relevant features of QFT on a classical FLRW
background. In section \ref{s3} we introduce the Hamiltonian set-up
to describe test fields on classical and quantum background
geometries and in section \ref{s4} we show how the two are related.
Section \ref{s5} contains a summary and presents the
outlook.\\

\emph{Remark:} Much of the detailed, recent work in LQC assumes that
the matter source is a massless scalar field $\T$ which, as we saw,
plays the role of a global, relational time variable. The overall
strategy is flexible enough to allow \emph{additional} matter
fields. The new issues that arise are technical, such as whether the
relevant operators continue to be essential self-adjoint. However
if, as in the simplest inflationary scenario, there is only a
massive scalar field ---and no massless ones--- one faces new
conceptual issues. In this case the scalar field serves as a good
time variable only `locally'. That is, one has to divide evolution
in `epochs' or `patches' in each of which the scalar field is
monotonic along dynamical trajectories. The discussion of the
quantum bounce is not much more complicated because the bounce
occurs in a single patch \cite{aps4}. The problem of joining
together these `patches' on the other hand is more complicated.
Although it can be managed in principle (see e.g. \cite{cr-time}),
at present it seems difficult to handle in practice.


\section{Background quantum geometry}
\label{s2}

LQC provides us a non-perturbative quantum theory of FLRW
cosmologies. Because it is based on a Hamiltonian treatment,
relation to the classical FLRW models was spelled out through
dynamical trajectories in the classical phase space \cite{aps3}.
In particular, the emphasis has been on the relational Dirac
observables, such as the matter density, anisotropies and
curvature at a given value of the scalar field. On the other hand,
quantum field theory on classical FLRW backgrounds is developed on
given classical space-times, rather than on dynamical trajectories
in the phase space of general relativity. Therefore, as a first
step we need to reformulate one of these descriptions using the
paradigm used in the other. In this section, we will recast the
LQC description, emphasizing space-times over phase space
trajectories. Relation to the cosmology literature will then
become more transparent.

We will focus on the k=0, $\Lambda=0$ FLRW models with a massless
scalar field as source. To avoid a discussion of boundary conditions
on test fields in section \ref{s3}, we will assume that the spatial
3-manifold is $\mathbb{T}^3$, a torus with coordinates $x^i \in (0,
\ell)$. It will be clear from our discussion that with appropriate
changes the analysis can be extended to include a cosmological
constant, or anisotropies, or closed k=1 universes.

\subsection{Space-time geometries and phase space trajectories}
\label{s2.1}

In this subsection we will clarify the relation between various
notions of time that feature in LQC and set up a dictionary between
the phase space and space-time descriptions.

Spatial homogeneity and isotropy implies that the space-time metric
has the form:
\be \label{g-proper} g_{ab}\,\dd x^a \dd x^b\, = \, -\dd t^2 +
q_{ij}\,\dd x^i \dd x^j \, \equiv \, -\dd t^2 + a^2 \dd {\xv}^2 \ee
where $q_{ij}$ is the physical spatial metric and $a$ is the scale
factor. Here the coordinate $t$ is the proper time along the world
lines of observers moving orthogonal to the homogeneous slices.

As explained in section \ref{s1}, in LQC one uses a relational time
defined by a massless scalar field which serves as a matter source.
Because of this and because we will also have a test scalar field
$\varphi$ in section \ref{s3}, we will denote the massless scalar
source by $T$. Since $T$ satisfies the wave equation with respect to
$g_{ab}$, in LQC it is most natural to consider the \emph{harmonic
time coordinate} $\tau$ satisfying $\Box \tau =0$. Then the
space-time metric assumes the form
\be \label{g-harmonic} g_{ab}\,\dd x^a \dd x^b \,\, =\,\, -a^6\, \dd
\tau^2 + q_{ij}\,\dd x^i \dd x^j \,\, \equiv\,\,  -a^6\, \dd\tau^2 +
a^2 \dd {\xv}^2 \ee

Let us now spell out the relation of this space-time metric to the
phase space trajectories.  In LQC, the gravitational part of the
phase space is conveniently coordinatized by a canonically
conjugate pair $(\v,b)$ where $\v$ is essentially the volume of
the universe and $b$, the Hubble parameter $\dot{a}/a$ (where, as
usual, the `dot' refers to derivative w.r.t. proper time $t$)
\cite{acs,cs1}. More precisely, the volume is given by

\be \label{V1} V \equiv \ell^3a^3\, =\, 2\pi \gamma \lp^2 |\v|\ee
and the Hubble parameter by $\dot{a}/a = b/\gamma$, where $\gamma$
is the so called Barbero-Immirzi parameter of LQG.%
\footnote{Following LQG, in LQC one uses orthonormal frames rather
than metrics. Since these frames can regarded as `square-roots' of
metrics, the configuration space is doubled. $\v, b \in \R^2$ are
constructed from the orthonormal frame and its time derivative,
and the sign of $\v$ depends on the orientation of the frame. The
canonical commutation relations are: $[\h{b}, \, \h{\v} ] = 2i$.}
(Its value, $\gamma \approx 0.24$, is fixed by black hole entropy
calculations.) \emph{Throughout this paper, we will pass freely
between $V, \v$ and the scale factor $a$.}

The canonically conjugate pair for the scalar field is $(\T, \pT)$.
Dynamics is generated by the Hamiltonian constraint, $NC$, where $N$
is the lapse function and $C$ the constraint function:
\label{C1} \be C = \f{\pT^2}{2V} - \f{3}{8\pi G}\,
\f{b^2}{\gamma^2}\,V \, \approx 0\ee
where, as usual, the weak equality holds on the constraint
hypersurface. If one uses the time coordinate $t$, then it follows
from (\ref{g-proper}) that the lapse is $N_t =1$, while if one uses
$\tau$ (\ref{g-harmonic}) implies that the lapse is $N_\tau = a^3$.
In the second case, the Hamiltonian constraint is:
\be \label{hc1} C_\tau := N_\tau C \,\equiv\, \f{\pT^2}{2\ell^3} -
\f{3}{8\pi G}\, \f{b^2}{\gamma^2}\, \f{V^2}{\ell^3}\, ,\ee
whence the time evolution of the scalar field is given by
\be \label{T} \T = \f{\pT}{\ell^3}\,\, \tau\ . \ee
(Here we have set the integration constant to zero for simplicity).
$\pT$ is a constant of motion which, for definiteness, will be
assumed to be positive. Then, as one would expect, in any solution
to the field equations the scalar field $\T$ grows linearly in the
harmonic time $\tau$ . Thus, although $\T$ does not have the
physical dimensions of time, it is a good evolution parameter.
Therefore, following the LQC literature, we will refer to it as the
\emph{relational time} parameter. On any given solution, we can
freely pass from $\tau$ to $T$ and write the space-time metric as:
\be \label{g-phi} g_{ab}\, \dd x^a \dd x^b\,\, =\,\, -
\f{a^6\ell^6}{\pT^2}\, \dd\T^2 + q_{ij}\, \dd x^i \dd x^j
\,\,\equiv\,\, \f{a^6\ell^6}{\pT^2}\, \dd\T^2 + a^2 \dd {\xv}^2 \ee
The only difference from (\ref{g-harmonic}) is that the lapse is
modified: $N_{\T} = (\ell^3/{\pT})\, N_{\tau}$ whence, in any
\emph{given} space-time, two lapse functions are related just by a
constant. However, in the \emph{phase space}, $\pT$ varies from one
dynamical trajectory to another, whence the relation is much more
subtle. If we regard $\T$ as a parameter, $\tau$ evolves
non-trivially on the full phase space, and vice versa. In quantum
gravity, we do not have a fixed space-time but a probability
amplitude for various geometries. Therefore, the situation in the
phase space is a better reflection of what happens in the quantum
theory. Indeed, as we will see in section \ref{s3}, the difference
between $\tau$ and $\T$ plays a deep role there.

Since the relation between the phase space and space-time notions is
important for our subsequent discussion, we will conclude with a
useful dictionary:
\begin{itemize}

\item A point in the phase space $\, \leftrightarrow\,$ A
    homogeneous slice in space-time (i.e., ${\mathbb T}^3\times
    \R$) equipped with the initial data for the gravitational
    and scalar field;
\item A curve in the phase space along which $\T$ is monotonic
    $\, \leftrightarrow\,$ A metric $g_{ab}$ and a scalar field
    $\T$ on space-time;
\item A curve in the phase space along which $\T$ is monotonic
    \emph{and} $\pT$ is constant $\, \leftrightarrow\,$ A metric
    $g_{ab}$ and a scalar field $\T$ satisfying $\Box T =0$ on
    space-time;\, and, finally,
\item  A dynamical trajectory in the phase space $\,
    \leftrightarrow\,$ A solution $(g_{ab}, \T)$ to the
    Einstein-Klein Gordon equation on space-time.

\end{itemize}

\subsection{Quantum FLRW space-times}
\label{s2.2}

In LQC one first constructs the quantum kinematics for the symmetry
reduced models by faithfully mimicking the unique kinematics of LQG,
selected by the requirement of background independence
\cite{lost,cf}. One then writes the quantum counterpart of the
Hamiltonian constraint (\ref{hc1}) as a self-adjoint operator on the
kinematical Hilbert space:
\be \label{hc2} \h{C}_\tau\, \Psi_o(\v,\T)\, =\, -
\f{\hbar^2}{2\ell^3}\, \Big(\partial_{\T}^2 +
\Theta\Big)\,\Psi_o(\v,\T)\, , \ee
where $\Theta$ turns out to be a difference operator in $\v$ given
by
\be \label{Theta} \Theta \Psi_o (\v, \T) = \f{3\pi G}{\l^2}\, \v \,
\Big[ (\v + 2\l)\Psi_o(\v+4\l) - 4\v \Psi_o(\v) + (\v
-2\l)\Psi_o(\v-4\v)\Big]\, . \ee
Here, $\lambda^2 = 4\sqrt{3}\pi\gamma\lp^2 $ is the smallest
non-zero eigenvalue of the LQG area operator (on states relevant to
homogeneity and isotropy) \cite{awe1,awe2,aa-badhonef} and we use
subscript (or superscript) $o$ to emphasizes that structures
developed in this section refer to what will serve as the
\emph{background} quantum geometry. Physical states must satisfy
\footnote{Recall from footnote 2 that $\v \rightarrow -\v$
corresponds just to change in the orientation of the orthonormal
frame which does not change the metric. Since the theory does not
involve any spinor fields, physics is insensitive to this
orientation. Therefore states must also satisfy $\Psi(\v,\T) =
\Psi(-\v,\T)$.}
\be \label{hc3} \h{C}_\tau \Psi_o(\v,\T) = 0 \, .\ee
A standard `group averaging procedure', which is applicable to a
wide class of constrained systems, then provides the scalar product
enabling us to construct the physical Hilbert space $\Hpo$. Since
the form of the constraint (\ref{hc2}) resembles the Klein Gordon
equation on a (fictitious) static space-time coordinatized by $\v,
\T$, as one might expect, $\Hpo$ is built out of `positive frequency
solutions' to (\ref{hc2}). More precisely, $\Hpo$ consists of
solutions to
 \be \label{hc4} -i\hbar \partial_\T \Psi_o (\v, \T) =  \h{H}_o
\Psi_o(\v,\T)\quad {\rm where} \quad \h{H}_o = \hbar\sqrt{\Theta}\,
.\ee
with finite norm with respect to the scalar product
\be\label{ip1} \bra \Psi_o,\,\Psi_o\ket = \f{\lambda}{\pi}\,
\sum_{\v = 4n\lambda}\, \f{1}{|\v|}\,\bar{\Psi}_o(\v, \T_0)\,
\Psi_o^\prime (\v, \T_0)\, . \ee
where the right side can be evaluated at any internal time $\T_0$.
Note that in their $\v$ dependence physical states have support on
the lattice $\v = 4n\lambda$, where $n$ ranges over all integers
(except zero). We will generally work in the Schr\"odinger
representation. Then, the states can be regarded as functions
$\Psi_o(\v)$ of $\v$ which have finite norm (\ref{ip1}) and which
evolve via (\ref{hc4}). The Hilbert space spanned by $\Psi(\v)$ will
be denoted by $\H_{\rm geo}$. For later use we note that the
classical expression (\ref{V1}) of volume implies that the volume
operator $\h{V}$ acts on $\H_{\rm geo}$ simply by multiplication:
\be \label{V2} \h{V}\Psi_o(\v) = 2\pi \gamma \lp^2 |\nu|
\Psi_o(\v)\, .\ee

Every element $\Psi_o$ of $\Hpo$ represents a 4-dimensional
quantum geometry. However, to make contact with QFT on classical
FLRW space-times, we are interested only in a subset of these
states which can be described as follows. Choose a classical,
expanding FLRW space-time in which $\pT \gg \hbar$ (in the
classical units $G$=$c$=1) and a homogeneous slice at a late time
$\T_o$, when the matter density and curvature are negligibly small
compared to the Planck scale. This defines a point $p$ in the
classical phase space. Then, one can introduce coherent states
$\Psi_o (\v, \T_o)$ in $\H_{\rm geo}$ which are sharply peaked at
$p$ \cite{aps2,aps3,acs}. Let us `evolve' them in the internal
time $\T$ using (\ref{hc4}). One can show \cite{aps3,acs} that
these states remain sharply peaked on the classical trajectory
passing through $p$ for all $\T > \T_o$. In the backward
time-evolution, it does so till the density reaches approximately
$1\%$ of the Planck density. As explained in section \ref{s1},
even in the deep Planck regime the wave function remains sharply
peaked but the peak now follows an effective trajectory which
undergoes a quantum bounce. At the bounce point the matter
density attains a maximum, $\rho_{\rm max} \approx 0.41 \rp$.%
\footnote{The existence of this maximum value does \emph{not} follow
simply from the fact that $|\v|$ is bounded below by $4\lambda$. Its
origin is more subtle \cite{acs,klp}: $\h\rho = (1/2)\, \h{V}^{-1}
\h{P}_{(\T)}^2 \h{V}^{-1}$ and the maximum value, $0.41\rp$, of
$\langle \hat\rho \rangle$ is the same no matter how large $\pT =
\langle\h{P}_{(\T)}\rangle$ is.}
After the bounce the density and the space-time curvature start
decreasing and once the density falls below about $1\%$ of the
Planck density, the effective trajectory becomes essentially
indistinguishable from a classical FLRW trajectory. Although the
effective trajectory cannot be approximated by any classical
solution in a neighborhood of the bounce point, $\pT$ is constant
along the entire effective trajectory. The dictionary given at the
end of section \ref{s2.1} then implies that the effective space-time
has a contracting FLRW branch in the past and an expanding FLRW
branch in the future. The scalar field $\T$ satisfies $\Box \T =0$
everywhere but Einstein's equations break down completely in an
intermediate region. Thanks to the quantum evolution equation
(\ref{hc3}), the two branches are joined in a deterministic fashion
in this region. \emph{By a quantum background geometry, we will mean
a physical state $\Psi_o(\v,\T)$ with these properties.} There is a
large class of such states and our considerations will apply to all
of them.

Of particular interest to us are the volume operators $\h{V}_{\T_0}$
on $\Hpo$ representing the volume of the universe at any fixed
instant $\T_0$ of internal time:
\be [\h{V}_{\T_0} \Psi_o](\v,\T)\, =\, e^{(i/\hbar)\h{H}_o
(\T-\T_0)}\,\, (2\pi \gamma\lp^2 |\v|)\,\, e^{-(i/\hbar)\h{H}_o
(\T-\T_0)}\,\, \Psi_o(\v, \T)\, . \ee
Thus, the action of $\h{V}_{\T_0}$ on any physical state $\Psi_o(\v,
\T)$ is obtained by evolving that state to $\T=\T_0$, acting on it
by the volume operator and then evolving the resulting function of
$\v$ using (\ref{hc4}). Each $\hat{V}_{\T_0}$ is a positive definite
self-adjoint operator. Hence one can define any (measurable)
function of $\h{V}_{\T_0}$ ---such as the scale factor
$\h{a}_{\T_0}$--- via its spectral decomposition. Finally, the
matter density operator $\h\rho_{\T_0}$ at time $\T=\T_0$ is given
by
\be \h\rho_{\T_0}\, =\, \f{1}{2}\,\,\h{V}^{-1}_{\T_0}\,
\h{P}_{(\T)}^2\, \h{V}^{-1}_{\T_0}\, \equiv\,
\f{\hbar^2}{2}\,\,\h{V}^{-1}_{\T_0}\, \Theta\, \h{V}^{-1}_{\T_0}\,
.\ee
As explained above, in background quantum geometries $\Psi_o(\v,\T)$
considered in this paper, the expectation values of $\h{\rho}_{\T}$
attain their maximum value $\rho_{\rm max} \approx 0.41 \rp$ at the
bounce point.

In the kinematical setting, $\h{\v}, \h{\T}, \h{P}_{(\T)}, \Theta$
are independent self-adjoint operators. However, in the passage to
the physical Hilbert space $\Hpo$ a `de-parametrization' occurs as
in the quantum theory of a parameterized particle (see, e.g.,
\cite{at}). On the physical sector \emph{we no longer have an
operator $\h\T$ but just a parameter $\T$} and the operator
$\h{P}^2_{(\T)}$ gets identified with $\hbar^2\Theta$. Consequently,
the space-time metric (\ref{g-phi}) can be represented as a
self-adjoint operator on $\Hpo$ as follows \cite{klp}:
\be \label{g-op} \h{g}_{ab}\, \dd x^a \dd x^b\,\, =\,\, - :
{\h{V}^2_{\T}}{\h{H}_o^{-2}}:\,\, \dd\T^2 + \h{q}_{ij}\, \dd x^i \dd
x^j \,\,\equiv\,\, \, :{\h{V}^2_{\T}}{\h{H}_o^{-2}}:\,\,\dd\T^2 +
\h{V}^{2/3}_{\T} \dd {\xv}^2 \,.\ee
Thus, the geometry is quantum because the metric coefficients
$\h{g}_{TT}$ and $\h{q}_{ij}$ are now quantum operators. In
(\ref{g-op}), a suitable factor ordering ---denoted by $:\,\,\, :$
--- has to be chosen because the volume operator $\h{V}_{\T}$ does
not commute with the Hamiltonian $\h{H}_o$ of the background quantum
theory. The simplest choice would be to use an anti-commutator but
it would be more desirable if the ordering is determined by some
general principles. (Note that $\h{H}_o^{-2}$ is well-defined
because $\h{H}_o$ is a positive self-adjoint operator.)

\section{The test quantum field}
\label{s3}

This section is divided in to two parts. In the first we summarize
the essential features of QFT on classical FLRW backgrounds in a
language that is well-suited for our generalization to quantum
backgrounds and in the second we carry out the generalization.

\subsection{QFT on classical FLRW backgrounds}
 \label{s3.1}

As in section \ref{s2}, let us fix a 4-manifold $M = \mathbb{T}^3
\times \R$, equipped with coordinates $x^j \in (0, \ell)$ and $x_0
\in \mathbb R$. Consider on it a FLRW 4-metric $g_{ab}$ given by
\be g_{ab} \dd x^a \dd x^b = -N_{x_0}^2(x_0) dx_0^2\ +\ a^{2}(x_0)
\dd {\xv}^2 \, ,\ee
where, as usual, the lapse function $N_{x_0}$ depends on the choice
of time coordinate $x_0$. Consider a real, massive, test Klein
Gordon field $\vp$ satisfying $(\Box - m^2)\,\vp =0$ on this
classical space-time $(M,g_{ab})$. Note that $\vp$ is \emph{not}
required to be homogeneous. Quantum theory of this field can be
described with various degrees of rigor and generality. As explained
in section \ref{s1}, in this paper, we will consider the simplest
version in terms of mode decomposition.

The canonically conjugate pair for the test scalar field consists of
fields $(\vp, \pphi)$ on a $x_0 = {\rm const}$ slice. Let us perform
Fourier transforms:
\be \vp(x_j, x_0) = \f{1}{(2\pi)^{3/2}} \sum_{{\kv}\in \L} \,
\vp_{\kv} (x_0)\, e^{i k_jx^j}\quad {\rm and} \quad \pphi(x_j, x_0)
= \f{1}{(2\pi)^{3/2}} \sum_{\kv\in \L} \, \pi_{\kv}(x_0)\, e^{i
k_jx^j}\, ,\ee
where $\L$ is the 3-dimensional lattice spanned by $(k_1,k_2,k_3)\in
((2\pi/\ell)\,\, \mathbb{Z})^3$,  $\Z$ being the set of integers.
The Fourier coefficients are canonically conjugate, $\{\vp_{\kv},\,
\pi_{\vec{k'}}\} = \delta_{\kv,\,-\vec{k'}}$ and, since
$\vp(\xv,x_0)$ is real, they satisfy the conditions: $\vp_{\kv} =
\bar{\vp}_{-\vec{k}}$ and $\pi_{\kv} = \bar{\pi}_{-\vec{k}}$. The
time dependent Hamiltonian (generating evolution in $x_0$) is given
by:
\ba H_\vp (x_0) &=& \f{1}{2}\, \int \f{N_{x_0}(x_0)}{a^3(x_0)}\,
\Big[\pphi^2 + a^4(x_0) (\partial_i\vp )^2 + m^2 a^6(x_0)\,
\vp^2 \Big]\, \dd^3x  \nonumber\\
&=& \f{N_{x_0}(x_0)}{2a^3(x_0)}\, \sum_{\kv\in \L}\, \bar{\pi}_{\kv}
\pi_{\kv} + ({\kv}^2 a^4(x_0) + a^6(x_0) m^2)\,\bar{\vp}_{\kv}
\vp_{\kv}\, . \ea

In the literature, the test scalar field $\vp$ is often regarded as
an assembly of harmonic oscillators, one for each mode. To pass to
this description, first note that because of the reality conditions,
the Fourier modes are inter-related. One can find an independent set
by, e.g., considering the sub-lattices $\L^\pm$ of $\L$ as follows:
\ba \L^+ &=& \{ {\kv}: k_3 >0\}\, \cup \{{\kv}: k_3 =0, k_2>0\} \cup
\{\kv: k_3 =0, k_2=0, k_1>0 \}\quad {\rm and}\nonumber\\
\L^- &=& \{\kv: -\vec{k} \in \L^+ \}\, .\ea
Then, for each $\kv \in \L^+$, we can introduce real variables
$q_{\pm\kv}, p_{\pm\kv}$,
\be \vp_{\kv} =\f{1}{\sqrt 2}( q_{\kv} + i q_{-\vec{k}}), \quad{\rm
and}\quad
     \pi_{\kv} = \f{1}{\sqrt 2} (p_\kv + i p_{-\vec{k}}). \ee
The pair  $(q_{\pm\kv},\, p_{\pm\kv})$ is canonically conjugate for
each $\kv \in \L^+$. In terms of these variables, the Hamiltonian
becomes
\be H_\vp(x_0) = \f{N_{x_0}(x_0)}{2a^3(x_0)}\, \sum_{\kv \in \L }\,
{p}^2_{\kv} +( {\kv}^2 a^4(x_0)  + m^2 a^6(x_0))\, {q}^2_{\kv}
\ee
where we have set $q_0:= \vp_{\vec{k}=0}$ and $\pi_0 :=
\pi_{\kv=0}$. Thus, the Hamiltonian for the test field is the same
as that for an assembly of harmonic oscillators, one for each $\kv
\in \L$.

To pass to the quantum theory, let us focus on just one mode
${\kv}$. Then we have a single harmonic oscillator. So the Hilbert
space is given by $H_{\kv} = L^2(\mathbb{R})$, the operator
$\h{q}_{\kv}$ acts by multiplication, $\h{q}_{\kv} \psi({q}_{\kv}) =
q_{\kv} \psi(q_{\kv})$ and $\h{p}_{\kv}$ acts by differentiation
$\h{p}_{\kv} \psi (q_{\kv}) = -i\hbar \dd \psi/\dd q_{\kv}$. The
time evolution is dictated by the time dependent Hamiltonian
operator $\h{H}_{\kv}(x_0)$:
\be \label{sch1} i\hbar \partial_{x_0}\psi(q_{\kv}, x_0)\, =\,
\h{H}_{\kv}(x_0) \psi(q_{\kv}, x_0)\, \equiv\,
\frac{N_{x_0}(x_0)}{2a^3(x_0)}\, \Big[\p_{\kv}^2 + ({\kv}^2
a^{4}(x_0)+ m^2a^6(x_0))\q_{\kv}^2 \Big]\, \psi(q_{\kv}, x_0). \ee

In this theory, there is considerable freedom in choosing the time
coordinate $x_0$ (and hence the lapse function $N_{x_0}$). One
generally chooses $x_0$ to be either the conformal time $\eta$ or
the proper time $t$. However, as we saw in section \ref{s2.2}, in
quantum geometry only the relational time $\T$ is a parameter;
$\eta, t$ and even the harmonic time $\tau$ become operators \cite{klp}.
Therefore, in QFT on a quantum geometry, while it is relatively
straightforward to analyze evolution with respect to $\T$,
conceptually and technically it is more subtle to describe evolution
with respect to conformal, proper or harmonic time (as it requires
the introduction of conditional probabilities). In the standard QFT
on classical FLRW space-times, on the other hand, $\T$ plays no
role; indeed, the source of the background geometry never enters the
discussion. This tension is conceptually significant and needs to be
resolved to relate QFT on classical and quantum FLRW geometries.

\subsection{QFT on quantum FLRW backgrounds}
\label{s3.2}

Recall first that in full general relativity dynamics is generated
by constraints. Our system of interest is general relativity coupled
to a massless scalar field $\T$ and a massive scalar field $\vp$,
where $\T$ is spatially homogeneous and $\vp$ is in general
inhomogeneous but regarded as a \emph{test} field propagating on the
homogeneous, isotropic geometry created by $\T$. Therefore, we can
start with the constraint functions on the \emph{full} phase space
of the gravitational field, $\T$ and $\vp$, but impose isotropy and
homogeneity on the gravitational field and $\T$ and retain terms
which are at most quadratic in $\vp$ and $\pphi$. The fact that we
are ignoring the back reaction of $\vp$ on the gravitational field
implies that, among the infinitely many constraints of this theory,
only the zero mode of the scalar constraint is relevant for us. That
is, we need to smear the scalar constraint \emph{only} with
homogeneous lapse functions (and can ignore the Gauss and the vector
constraints). For concreteness, as in section \ref{s2.1}, we will
choose the harmonic time coordinate $\tau$ and the corresponding
lapse function $N_\tau = a^3$. Then, in the truncated theory now
under consideration, the scalar constraint (\ref{C1}) is replaced
by:
\be \label{C2}  C_\tau \equiv N_\tau C \,=\, \f{\pT^2}{2\ell^3} -
\f{3}{8\pi G}\, \f{b^2}{\gamma^2}\,\f{V}{\ell^3} \, + \f{1}{2}\,\int
[\pphi^2+ a^4 (\partial_i \vp)^2 + m^2 a^6 \vp^2]\, \dd^3x \approx\,
0 \ee
(Recall that the volume and the scale factor are related by $V=
\ell^3a^3$.) If we focus just on the $\kv$th mode, the constraint
simplifies further:
\be \label{C3}  C_{\tau, \kv}\, = \, \f{\pT^2}{2\ell^3} - \f{3}{8\pi
G}\, \f{b^2}{\gamma^2}\,\f{V}{\ell^3} \, + H_{\tau,\kv}\ee
where
\be H_{\tau,\kv} = \f{1}{2}\, [\,{p}_{\kv}^2 +  ({\kv}^2 a^4 + m^2
a^6) q_{\kv}^2\,]\ee

In quantum theory, then, physical states $\Psi(\v, q_\kv,\T)$ must
be annihilated by this constraint, i.e., must satisfy:
\be \label{hc5} -\hbar^2\partial_\T^2\, \Psi(\v,q_\kv,\T)\, =\, [\,
\h{H}_o^2 - 2\ell^3\, \h{H}_{\tau,\kv}\,]\, \Psi(\v,q_\kv,\T)\, ,
\ee
where as in section \ref{s2.2}, $\h{H}_o^2 = \hbar^2\Theta$ is the
difference operator defined in (\ref{Theta}). (Although $\h{a}$ is
an operator, it commutes with $\h{q}_{\kv}$ and $\h{p}_{\kv}$ on the
kinematical Hilbert space. So there are no factor ordering
subtleties in the definition of $\h{H}_{\tau,\kv}$.) As in section
\ref{s2.2}, the construction of the physical inner product requires
us to take the `positive-frequency' square-root of this equation.
More precisely, on the tensor product $\H_{\rm geo}\otimes L^2(\R)$
of the quantum geometry Hilbert space $\H_{\rm geo}$ and the
$\kv$-mode Hilbert space $L^2(\R)$, the operator $[\,\h{H}_o^2 -
2\ell^3\, \h{H}_{\tau,\kv}\,]$ on the right hand side of (\ref{hc5})
is symmetric and we assume that it can be made self-adjoint on a
suitable domain. On the physical Hilbert space, this operator gets
identified with $\,\hat{P}^2_{(\T)}$. Since classically $\pT^2$ is a
positive Dirac observable, we are led to restrict ourselves to the
positive part of the spectrum of $[\,\h{H}_o^2 - 2\ell^3\,
\h{H}_{\tau,\kv}\,]$ and then solve the evolution equation
\be \label{hc6} -i\hbar\, \partial_T\, \Psi(\v,q_\kv,\T) =
[\h{H}_o^2 - 2\ell^3\, \h{H}_{\tau,\kv}]^{\f{1}{2}}\,
\Psi(\v,q_\kv,\T)\,=:  \h{H} \Psi(v, q_\kv, T) .\ee
The solutions are in the physical Hilbert space $\Hp$ of the
truncated theory provided they have a finite norm with respect to
the inner product:
\be\label{ip2} \bra \Psi_1 |\Psi_2\ket = \f{\lambda}{\pi}\, \sum_{\v
= 4n\lambda}\, \f{1}{|\v|}\,\int_{-\infty}^{\infty} \dd q_{\kv}\,\,
\bar{\Psi}_1(\v,q_\kv, \T_0)\, \Psi_2 (\v,q_\kv,\T_0)\,  \ee
where the right side is evaluated at \emph{any} fixed instant of
internal time $\T_0$. As one might expect, the physical observables
of this theory are the Dirac observables of the background geometry
---such as the time dependent density and volume operators
$\h{\r}(\T)$ and $\h{V}(\T)$--- and observables associated with the
test field, such as the mode operators $\h{q}_{\kv}$ and
$\h{p}_{\kv}$.

Formally, this completes the specification of the quantum theory of
the test field $\h{\phi}$ on a quantum FLRW background geometry. We
have presented this theory (as well as the QFT a classical
background in section \ref{s3.1}) using the Schr\"odinger picture
because this is the description one is naturally led to when,
following Dirac, one imposes quantum constraints to select physical
states. However, at the end of the process it is straightforward to
re-express the theory in the Heisenberg picture.\\ 

\emph{Remark:} In this section we began with the constraint
(\ref{C2}) on the classical phase space spanned by $(\v,b; \T, \pT;
\vp, \pphi)$. Solutions to this theory do include a back reaction of
the field $\vp$ but just on the homogeneous mode of the classical
geometry. In the final quantum theory, the Hamiltonian of the field
$\vp$ features on the right side of (\ref{hc6}) whence, in the
Heisenberg picture, it affects the evolution of geometric operators.
As in the classical theory, this evolution incorporates back
reaction of the field $\h\vp$ but just on the homogeneous mode of
the quantum geometry. Mathematically, we have a closed system
involving $\h\v, \h\vp,\T$ whence this inclusion of the back
reaction is consistent. However, physically it is not as meaningful
because we have ignored the back reaction at the same order that
would add inhomogeneities to the quantum geometry. So, from a
physical viewpoint, \emph{all} corrections to quantum geometry which
are quadratic in $\h\vp$ should be consistently ignored. We will
explicitly impose this restriction in section \ref{s4.3}. However,
the classical theory determined by (\ref{C2}) and the quantum theory
constructed in this section can be directly useful in some
applications where it is meaningful to ignore inhomogeneous metric
perturbations and study the homogeneous mode, including the back
reaction corrections.

\section{Comparison}
\label{s4}

In this section we will compare QFT on a classical background
discussed in section \ref{s3.1} and QFT on quantum FLRW geometries
discussed in section \ref{s3.2}. The discussion is divided into
three sub-sections which provide the successively stronger
simplifications of the dynamical equation (\ref{hc6}) that are
needed to arrive at the dynamical equation (\ref{sch1}) on a
classical FLRW space-time.

\subsection{Simplification of the evolution equation}
\label{s4.1}

Let us begin by using the test field approximation. Since the back
reaction of the scalar field $\vp$ is neglected, the theory
constructed in section \ref{s3.2} can be physically trusted only on
the sector on which $\h{H}_o^2$ dominates over ${2\ell^3}\,
\h{H}_{\tau,\kv}$. On this sector, one can expand out the
square-root on the right side of (\ref{hc6}) in a useful fashion.
First let us consider the regime in which the support of $\Psi
(\nu)$ is on $\nu \gg \lambda$. (For semi-classical states of
quantum geometry under consideration, this condition is not a real
restriction.) Furthermore, suppose for a moment that there is a
negative cosmological constant, i.e., $\h{H}^2_o$ is replaced by
$\h{H}_\Lambda^2 = \h{H}_o^2 + C \Lambda \nu^2$, where $C$ is a
positive constant. Then, one can show that Eq. (\ref{hc6}) modified by the
presence of a negative $\Lambda$ can be approximated by:
\be \label{expansion} -i\hbar\partial_\T\, \Psi(\v,q_\kv,\T) =
\Big(\h{H}_\Lambda \,-\, \big(\ell^{-3}
\h{H}_\Lambda\big)^{-\f{1}{2}}\,\, \h{H}_{\tau,\kv}\,\,
\big(\ell^{-3}\h{H}_\Lambda\big)^{-\f{1}{2}}\Big)\,
\Psi(\v,q_\kv,\T)\, .\ee
We will assume that the same approximation holds in the
$\Lambda=0$ case,%
\footnote{Thus, introduction of $\Lambda$ at this intermediate stage
is like a `regularization'. Alternatively, one can restrict oneself
to the case where there is a negative cosmological constant from the
beginning.}
i.e., we will use the following simplification of (\ref{hc6}):
%
%
%
%
%
\be \label{hc7} -i\hbar\partial_\T\, \Psi(\v,q_\kv,\T) =
\Big(\h{H}_o \,-\, \big(\ell^{-3} \h{H}_o\big)^{-\f{1}{2}}\,\,
\h{H}_{\tau,\kv}\,\, \big(\ell^{-3}\h{H}_o\big)^{-\f{1}{2}}\Big)\,
\Psi(\v,q_\kv,\T)\, .\ee
%

%
We will now show that the second term on the right side of
(\ref{hc7}) has a direct interpretation. In the classical theory,
$H_{\tau, \kv}$ is the Hamiltonian generating evolution in harmonic
time $\tau$. Since the corresponding lapse function $N_\tau$ is
related to the lapse function $N_\T$ corresponding to the relational
time $\T$ via $N_T = (P_\T \ell^3)^{-1}N_\tau $, the Hamiltonian
generating evolution in $\T$ is given by $H_{\T, \kv} =
(\ell^{-3}P_\T)^{-1} H_{\tau, \kv} \approx (\ell^{-3}H_o)^{-1}
H_{\tau, \kv}$, where in the last step we have again used the test
field approximation. The second term on the right side of
(\ref{hc7}) is \emph{precisely} a specific quantization of $H_{\T,
\kv}$. This is just as one would physically expect because the left
side of (\ref{hc7}) is the derivative of the quantum state with
respect to $\T$. Thus, we can rewrite (\ref{hc7}) as:
\be \label{hc8} -i\hbar\partial_\T\, \Psi(\v,q_\kv,\T) =
\big(\h{H}_o \,-\, \h{H}_{\T,\kv}\big)\,\Psi(\v,q_\kv,\T) \, .\ee
The non-triviality lies in the fact that this evolution equation
arose from a systematic quantization of the $(\v, \vp, \T)$-system
where geometry is also quantum. As in LQC we began with the quantum
constraint operator associated with the harmonic time, and then used
the group averaging procedure to find the physical Hilbert space.
This naturally led us to take a square root of the quantum
constraint and then a simplification which is valid in the test
field approximation automatically provided the extra factor to
rescale the lapse operator just in the right manner to pass from the
harmonic to the relational time. Thus, there is coherence between
the constrained dynamics, various notions of time involved,
deparametrization of the full theory and the test field
approximation.

\subsection{Interaction picture}
\label{s4.2}

The simplified evolution equation (\ref{hc8}) is rather analogous to
the Schr\"odinger equation (\ref{sch1}) in QFT on a classical FLRW
background. However, there are two key differences. First, in
(\ref{hc8}) the background geometry appears through \emph{operators}
$\h{V}$ and $\h{H}_o$ while in (\ref{sch1}) it appears through the
\emph{classical} scale factor $a(x_0)$ and (if we set $x_0 =\T$) the
constant $\ell^3/\pT = N_{\T}/a^3$ determined by the momentum of
background scalar field. The fact that there are operators on the
Hilbert space $\H_{\rm geo}$ of quantum geometry in one case and
classical fields on space-time $M$ in the second is not surprising.
But there is also a more subtle, second difference. The operators
$\h{H}_o$ and $\h{V}$ which features on the right side of
(\ref{hc8}) do \emph{not} depend on time:%
\footnote{This also occurs in the classical theory. There, in place
of the Hamiltonian, we have the constraint function $C =
{\pT^2}/{2V} - ({3}/{8\pi G})\, ({b^2V}/{\gamma^2})$ on the phase
space. $b,V$ which appear in the expression are determined just by
the point at which $C$ is evaluated; there is no time parameter on
which they could depend! This is in fact the origin of the fact the
$\h{V}$ and $\h{H}_o$ in (\ref{hc8}) do not depend on time.}
$\h{V}\,\Psi(\v,q_\kv,\T) = 2\pi \gamma \lp^2 |\v|\,
\Psi(\v,q_\kv,\T)$ and $\h{H}_o\, \Psi(\v,q_\kv,\T) = \hbar
\sqrt{\Theta}\, \Psi(\v,q_\kv,\T)$. The scale factor $a(x_0)$ that
appears in (\ref{sch1}) on the other hand is explicitly time
dependent. This is because while (\ref{hc7}) provides a quantum
evolution equation for the state $\Psi(\v,q_\kv,\T)$ that depends on
(the $\kv$th mode of) the test field $\vp$ \emph{and} the quantum
geometry (encoded in $\v$), (\ref{sch1}) evolves the state
$\psi(q_\kv,\T)$ just of the test scalar field on the given time
dependent background geometry (encoded in $a(x_0)$).

To make the two evolutions comparable, therefore, we need to recast
(\ref{hc8}) in such a way that the test field evolves on a
background, \emph{time-dependent} quantum geometry. This can be
readily achieved by working in the `interaction picture'. More
precisely, it is natural to regard $\h{H}_o$ in (\ref{hc8}) as the
Hamiltonian of the heavy degree of freedom and $\h{H}_{\T,\kv}$ as a
perturbation governing the light degree of freedom and, as in the
interaction picture, set
\be \Psi_\inter (\v,q_\kv,\T) := e^{-(i/\hbar) \h{H}_o\,
(\T-\T_0)}\,\Psi (\v,q_\kv,\T)\, .\ee
where $\T_0$ is any fixed instant of relational time. Then,
(\ref{hc8}) yields the following evolution equation for
$\Psi_\inter$:
\ba \label{hc9} i\hbar \partial_\T\, \Psi_\inter(\v,q_\kv,\T) &=&
\f{1}{2}\, \big(\ell^3 \h{H}_o\big)^{-\f{1}{2}}\, \Big[ p^2_\kv +
({\kv}^2 \h{a}^4(\T)\, +\, m^2 \h{a}^6(\T)) q_{\kv}^2 \Big]
\big(\ell^3
\h{H}_o\big)^{-\f{1}{2}}\, \Psi_\inter(\v,q_\kv,\T)\nonumber\\
&=:& \h{H}^\inter_{\T,\kv}\,\, \Psi_\inter(\v,q_\kv,\T)\, .
 \ea
Here the operators $\h{a}(T)$ (and their powers) are defined on the
Hilbert space $\H_{\rm geo}$ of quantum geometry (now tied to the
internal time $\T_o$):
\be \h{a}(\T) = e^{-(i/\hbar)\, \h{H}_o (\T-\T_o)}\,\, \h{a}\,\,
e^{(i/\hbar)\, \h{H}_o (\T-\T_o)}\, \quad{\rm with}\quad \hat{a}\ =\
\frac{1}{\ell}|\hat{V}|^{\frac{1}{3}} . \ee

Thus, in this `interaction picture' quantum geometry is in effect
described in the Heisenberg picture ---states of quantum geometry
are `frozen' at time $\T= \T_o$ but the scale factor operators
evolve--- while the test field is described using the Schr\"odinger
picture. Therefore, the quantum evolution equation (\ref{hc9}) is
now even more similar to the Schr\"odinger equation (\ref{sch1}) for
the test field on a classical background. However, the lapse
$\h{N}_\T$ and powers of the scale factor $\h{a}$ are still
operators on $\H_{\rm geo}$. In the next subsection we will specify
the approximations necessary to reduce (\ref{hc9}) to (\ref{sch1}).

\subsection{Replacing geometric operators by their mean values}
\label{s4.3}

Let us now assume that the state $\Psi_\inter(\v,q_\kv,\T)$
factorizes as $\Psi_\inter(\v,q_\kv,\T) = \Psi_o(\v,\T_0) \otimes
\psi(q_\kv,\T)$ where $\Psi_o(\v,\T_0)$ is a quantum geometry state
introduced in section \ref{s2.2}, peaked at an effective LQC
geometry of the $(\nu,\vp)$-system. This assumption is justified
because $\vp$ is a test field, i.e., its back reaction is ignored.
Then, (\ref{hc9}) further simplifies as follows
\ba \label{hc10} \Psi_o(\v,\T_0) \otimes [i\hbar \partial_{\T}\,
\psi(q_\kv, \T)] =\,\, \f{1}{2}\,\big[ (\ell^{-3} \h{H}_o)^{-1}\,
\Psi_o(\v, \T_0)\big] \, &\otimes&\, \big[\h{p}_{\kv}^2\,
\psi(q_\kv,\T)\big]\nonumber \\
+\, \f{1}{2}\,\big[ (\ell^{-3} \h{H}_o)^{-\f{1}{2}}\, ({\kv}^2
\h{a}^4(\T) + m^2 \h{a}^6(\T))\, (\ell^{-3} \h{H}_o)^{-\f{1}{2}}\,
\Psi_o(\v,\T_0)\big]\,&\otimes&\, \big[\h{q}_{\kv}^2\, \psi(q_\kv,
\T)\big]\ea

Let us now suppose that $\Psi_o(\v, \T_o)$ is normalized and take
the scalar product of (\ref{hc10}) with $\Psi_o(\v,\T_0)$. Then, we
obtain:
\ba \label{sch2} i\hbar\partial_\T\, \psi(q_\kv, \T)
 &=&\f{1}{2}\, \big{\langle}\, (\ell^{-3}\h{H}_o)^{-1}\big{\rangle}
 \,\, \h{p}_{\kv}^2\,\, \psi(q_\kv, \T)
 \,+\, \f{1}{2}\, \Big[\, {\kv}^2\, \big{\langle}\,
(\ell^{-3}\h{H}_o)^{-\f{1}{2}} \h{a}^4(\T)
(\ell^{-3}\h{H}_o)^{-\f{1}{2}} \big{\rangle}\nonumber\\
 &+& m^2 \, \big{\langle}\, (\ell^{-3}\h{H}_o)^{-\f{1}{2}}
 \h{a}^6(\T)
(\ell^{-3}\h{H}_o)^{-\f{1}{2}} \big{\rangle}\Big]\, \h{q}_{\kv}^2\,
\psi(q_\kv, \T)\, \ea
where $\langle\h{A}{\rangle}$ denotes the expectation value of the
operator $\h{A}$ in the quantum geometry state $\Psi_0$. Thus, in
this equation all geometrical quantities are c-numbers. Nonetheless,
(\ref{sch2}) \emph{is in general different from} (\ref{sch1})
\emph{because expectation values of products of operators do not
equal products of expectation values of operators.} We discuss the
differences and analogies below.

Eq (\ref{sch2}) tells us how the quantum state of the mode $q_\kv$
`evolves', but the background geometry is neither classical nor
quantum in the sense of section \ref{s2.2}. The mode knows about the
background geometry only through the three expectation values that
feature on the right side of (\ref{sch2}). Therefore one is led to
ask if there is an \emph{effective} classical FLRW space-time such
that the Schr\"odinger equation (\ref{sch1}) on it is equivalent to
(\ref{sch2}).

To address this question, let us begin with the plausible assumption
that the quantum geometry state $\Psi_0$ is sharply peaked at the
expectation values $\bar{P}_{(\T)}$ and $\bar{a}$ of $\h{H}$ and
$\h{a}$ respectively and, \emph{furthermore}, work in the
approximation in which quantum fluctuations of geometry can be
ignored. A priori this is a very strong simplification but, for
cosmological applications, this approximation can be justified
because the quantum geometries $\Psi_o(\v,\T)$ have incredibly small
dispersions along the entire effective trajectory \cite{cs2}. Then,
(\ref{sch2}) reduces to:
\be \label{sch3} i\hbar\partial_\T\, \psi(q_\kv, \T) =
\f{\bar{N}_\T}{2\bar{a}^3}\, \Big[\, \p_{\kv}^2 + ({\kv}^2
\bar{a}^{4}(x_0)+ m^2 \bar{a}^6(x_0)) \q_{\kv}^2\, \Big]\,
\psi(q_{\kv}, x_0)\, . \ee
This is exactly the Schr\"odinger equation (\ref{sch1}) governing
the dynamics of the test quantum field on a classical space-time
with scale factor $\bar{a}$ containing a massless scalar field $\T$
with momentum $\bar{P}_{(\T)} = \bar{a}^2\ell^3/\bar{N}_\T$. This is
the precise sense in which the dynamics of a test quantum field on a
classical background emerges from a more complete QFT on quantum
FLRW backgrounds. Note however that, even with this strong
simplification, the classical space-time is \emph{not} a FLRW
solution of the Einstein-Klein-Gordon equation. Rather, it is the
effective space-time $(M, \bar{g}_{ab})$ a la LQC on which the
quantum geometry $\Psi_o(\v,\T)$ is sharply peaked. But as discussed
in sections \ref{s1} and \ref{s2.1}, away from the Planck regime,
$(M, \bar{g}_{ab})$ is extremely well-approximated by a classical
FLRW space-time $(M, g^o_{ab})$. Thus, starting from quantum
geometry and making a series of well-motivated approximation, we
have arrived at a QFT of a test field $\vp$ which is a non-trivial
extension of the QFT on a standard $(M, g^o_{ab})$. It has the same
structure as the standard theory but is defined on a much larger
space-time in which the big bang is replaced by a quantum bounce and
there is an infinite pre-big-branch. Therefore, although the theory
developed in this section describes a test quantum field $\h{\vp}$
on classical backgrounds and approximates the standard QFT on
classical FLRW geometries at late times, it also contains a lot of
new physics, particularly in the Planck regime around the bounce.

Next, it is interesting to return to the equation (\ref{sch2}) and
\emph{not} make additional simplifications. One can still ask if
there is a classical metric tensor
 \be g^\prime_{ab}dx^adx^b\ =\ -{N^\prime}^2(\T)\, \dd\T^2 +
 {a^\prime}^2(\T)\, \dd\vec{x}^2\ee
such that  (\ref{sch2}) agrees with the Schr\"odinger equation
(\ref{sch1}) on $(M,g^\prime_{ab})$. For this agreement to hold, the
scale factor $a^\prime(T)$ and the lapse function $N^\prime(T)$
should satisfy the following system of equations:
\begin{align} N^\prime(T)\ &=\ \ell^3 {a^\prime}^3(T)\big{\langle}\,
\h{H}_o^{-1}\big{\rangle}\\
N^\prime(T)a^\prime(T)\ &=\ \ell^3 \big{\langle}\, \h{H}_o^{-\f{1}{2}}
\h{a}^4(\T) \h{H}_o^{-\f{1}{2}} \big{\rangle}\\
m^2\,N^\prime(T){a^\prime}^3(T)\ &=\ m^2\,\ell^3 \big{\langle}\,
\h{H}_o^{-\f{1}{2}} \h{a}^6(\T)\ell^{-3}\h{H}_o^{-\f{1}{2}}
\big{\rangle}.
 \end{align}
In the case then the test field is massless, the third equation
disappears and there is clearly a solution
$(N^\prime(T),\,a^\prime(T))$. But note that the interpretation of
(\ref{sch2}) as the evolution equation for $\psi(q_\kv, \T)$ on the
classical space-time $(M, g^\prime_{ab})$ is not entirely
satisfactory because, if the quantum geometry state is sharply
peaked at $\langle\h{a}\rangle = \bar{a}$ and $\langle
\h{P}_{(\T)}\rangle = \bar{P}_{(\T)}$, then $a^\prime(\T)\,\not=
\bar{a}(\T)$ and $N^\prime(T)\, \not=\,
\ell^3\,{\bar{a}^3}/{\bar{P}_{(\T)}}$. Thus, deductions about the
quantum geometry made from the dynamics of the test scalar field
would be different from those made by observing the geometry
directly, e.g., from the measurement of the Hubble parameter or of
the volume at the bounce point. Finally, in the case when the test
scalar field $\vp$ has mass, on the other hand, if the quantum
geometry fluctuations are not negligible, dynamics of the test field
given by (\ref{sch2}) cannot be interpreted as dynamics of the test
field on \emph{any} classical FLRW background.

\section{Discussion}
\label{s5}

Consider QFT of a massive, test, scalar field $\h\vp$ on a classical
FLRW space-time $(M, g_{ab}^o)$ with a massless scalar field $\T$ as
its matter source. Our main goal was to derive this theory from that
the scalar field $\h\vp$ on a quantum geometry $\Psi_o(\v,\T)$ that
replace $(M, g_{ab}^o)$ in LQC. Conceptually the two theories are
quite distinct:
\begin{itemize}
\item They use very different notions of time. In particular,
    the conformal time $\eta$ and the proper time $t$ used in
    the first are non-trivial operators in the second
\cite{klp};
\item In the first, dynamics is generated by a Hamiltonian while
    in the second, it has to be teased out of a constraint;
\item In the first, there is a fixed classical metric $g^o_{ab}$
    in the background which is used repeatedly in the
    construction of the QFT, while in the second there is only a
    probability distribution for various metrics encoded in
    $\Psi_o(\v,\T)$; and
\item While in the first theory the scale factor $a$
    is a given function on $M$, in the second theory we are
    confronted with quantum fluctuations of (different powers
    of) the operator $\h{a}$.
\end{itemize}
Our first task was to set up an appropriate framework to explore the
relation between the two theories in detail

To construct the second of these theories, in section \ref{s3} we
began with the constrained quantum system for the gravitational
field coupled with the scalar fields $\T$ and $\vp$ but made
simplifications to encode the idea that the space-time geometry and
$\T$ are homogeneous and $\vp$ is (inhomogeneous but) a test field
whose back reaction is ignored. This theory was de-parameterized by
singling out $\T$ as the relational time variable with respect to
which the gravitational field and $\vp$ evolve. The states of the
coupled system are then functions $\Psi(\v,\vp,\T)$ of volume $\v$
(or, equivalently, the scale factor) of the universe, the massive
test field $\vp$ and the massless scalar field $\T$. We found that
their inner product is given by (\ref{ip2}) and their dynamics is
governed by the Schr\"odinger equation (\ref{hc6}). Thus, a quantum
theory of the test field $\vp$ on quantum geometries could be
constructed although we do not have a fixed classical metric or a
fixed causal structure in the background.

In section \ref{s4} we made successive approximations to simplify
(\ref{hc6}), all of which are well-motivated by the set-up of the
problem:
\begin{itemize}
\item We regarded variables $(\v,\T)$ which provide the
    background geometry as the heavy degree of freedom and the
    test field $\vp$ as the light degree to simplify the
Hamiltonian operator in (\ref{hc6});
\item We assumed that the state $\Psi(\v,\vp,\T)$ can be
    expanded as $\Psi(\v,\vp,\T) = \Psi_o(\v, \T)\otimes
    \psi(\vp, \T)$ where $\Psi_o(\v,\T)$ is the quantum geometry
    that replaces the classical FLRW space-time in LQC, and took
    the scalar product of the evolution equation (\ref{hc6})
    w.r.t. the quantum geometry state $\Psi_o(\v,\T)$ to obtain
    an evolution equation for $\psi(\vp)$.
\item To simplify this equation on $\psi(\vp)$, we ignored the
    quantum fluctuations of geometry by replacing the
    expectation values of products of geometrical operators by
    products of their expectation values. The result was the
    standard Schr\"odinger equation (\ref{sch3}) for a test
    field $\vp$ on a classical background.
\end{itemize}

However, equation (\ref{sch3}) has two non-standard features. First,
the classical background is \emph{not} a FLRW space-time
$(M,g^o_{ab})$ but rather an effective space-time $(M,
\bar{g}_{ab})$ on which the LQC state $\Psi_o(\v,\T)$ is sharply
peaked. Second, the Schr\"odinger equation naturally arises with
$\T$ as the time variable. This is unusual from the perspective of
QFT on classical backgrounds because $\T$ is the massless scalar
field that acts as the source of the gravitational field while QFT
on classical backgrounds, as normally formulated, has no knowledge
of the source. Rather, the time variables that are normally used
---the conformal time $\eta$ or the proper time $t$--- arise
directly from the metric $g^o_{ab}$. However, from the perspective
of quantum geometry, these are unnatural because while $\T$ is a
parameter in that theory, as we noted above, $\eta$ and $t$ are not;
they get promoted to operators. Of course, once we have arrived at
the `lower' theory ---i.e., QFT on the classical space-time $(M,
\bar{g}_{ab})$--- it is straightforward to reformulate dynamics in
terms of either $\eta$ or $t$. But at a more fundamental level,
\emph{it is the relational time $\T$ that appears to be the natural
time parameter.} Finally, let us return to the first difference. The
effective space-time $(M, \bar{g}_{ab})$ is a non-trivial extension
of the FLRW solution $(M, g^o_{ab})$ in which the big bang is
replaced by a bounce and there is an infinite pre-big-bang branch.
However, FLRW solutions $(M, {g}^o_{ab})$ are excellent
approximations to the effective space-times $(M,\bar{g}_{ab})$ in
the expanding, post-big-bang branch \emph{away from the Planck
regime.} Furthermore, our QFT on effective space-times does reduce
to the standard one on FLRW solutions when the space-time curvature
is smaller than the Planck scale. Moreover, it provides a physically
interesting extension near and to the past of the big bounce.
Because $(M,\bar{g}_{ab})$ is non-singular, this theory opens a new
window on the Planck scale physics which was inaccessible to QFT on
classical FLRW solutions.

Thus, in this paper we have laid down foundations for further work
with applications to cosmology as well as mathematical physics. We
will conclude by indicating directions that are being currently
pursued. First, we need to include the back reaction of $\vp$ on
geometry, treating it as a perturbation. As far as the homogeneous
mode of the gravitational field is concerned, this is already
achieved in the evolution equation (\ref{hc6}) (see the remark at
the end of section \ref{s3.2}). Inclusion of inhomogeneous
gravitational perturbations remains an open issue. Second, we have
to analyze the quantum dynamics of the gauge invariant combinations
$\Phi$ of $\vp$ and the scalar perturbations of the metric. Here the
important step is to construct the Mukhanov variable $\Phi$ starting
from the full quantum constraint. Existing literature on
cosmological perturbations in the LQG setting \cite{dt,ghtw} is
likely to be directly useful in this task. The mathematical theory
of propagation of $\Phi$ on the quantum background geometry
$\Psi_o(\v,\T)$ would be rather similar to that of $\vp$ analyzed in
this paper. Third, we have to account for the origin of the massless
scalar field $\T$ which plays the role of time for us. It seems most
natural to have a single scalar field $\Phi$, the homogeneous mode
of which would provide the relational time parameter $\T$ and the
inhomogeneous modes, the physical perturbations that lead to
structure formation. This seems feasible. However, it is likely that
the resulting relational time will not be global. Thus, as remarked
at the end of section \ref{s1}, the analysis in quantum geometry may
have to be divided into `epochs' in each of which the homogeneous
part of $\Phi$ will serve as a relational time variable. If these
three steps can be carried out to completion, we will have a
coherent framework to analyze cosmological perturbations and
structure formation which is free from the limitations of a big bang
singularity. In particular, one will then be able to evolve
perturbations across the big bounce and study phenomenological
implications. Immediately after the big bounce, there is a short
epoch of super-inflation in LQC (see
\cite{mb2} and especially \cite{ps}). The possibility that
ramifications of this sudden and very rapid expansion may be
observable has drawn considerable attention of cosmologists
recently. A more complete QFT on quantum geometries will provide a
systematic avenue to analyze these issues.

The second direction for further work is motivated by mathematical
physics (although it too has some implications to cosmology). In
this paper we focused on a single mode of the scalar field $\vp$.
Inclusion of a finite number of modes is completely straightforward.
Inclusion of all modes, on the other hand, involves functional
analytic subtleties. Recall, however, that in quantum geometry, the
volume operator has a non-zero minimum value, $2\pi \gamma \lp^2
|\v|_{\rm min} = 8\pi\gamma\lambda \lp^2$. Therefore, in a certain
sense there is a built-in ultra-violet cut-off. A careful
examination may well reveal that this cut-off descends to the test
scalar field $\vp$, in which case $\vp$ would have only a finite
number of modes and the treatment presented here will suffice.
However, if this possibility is not realized, one would have to
resolve the functional analytical difficulties. Our first task is to
address these issues. Second, a number of ideas related to the
algebraic approach are being explored. This approach can be applied
directly to the effective space-times $(M, \bar{g}_{ab})$ that
emerge from LQC. What can one say about the (regularized)
stress-tensor of $\vp$ and its back reaction on the geometry? Is
there a sense in which the Schr\"odinger equation (\ref{hc6})
already includes these effects? More importantly, can one extend the
algebraic approach systematically to cosmological \emph{quantum}
geometries? At first the extension seems very difficult, if not
impossible, because so many of the structures normally used in the
algebraic approach to QFT on classical space-times use the fact that
we have access to a \emph{fixed} space-time metric. However, in the
cosmological context, additional structures ---such as a preferred
foliation--- naturally become available and they enable one to
construct the required $\star$-algebras of field operators in the
canonical setting. Also, the background quantum geometries
$\Psi_o(\v,\T)$ are rather well-controlled and one may be able to
use the fact that they are extremely sharply peaked around effective
space-times \cite{cs2}. Can one exploit this setting to introduce
the analogs of Hadamard states? We believe that such generalizations
are now within reach.

\section*{Acknowledgments} We have profited from discussions with
Alejandro Corichi, Klaus Fredenhagen, Tomasz Pawlowski and Param
Singh. This work was supported in part by the NSF grants
PHY0456913 and  PHY0854743  the Polish Ministerstwo Nauki i
Szkolnictwa Wyzszego grants 1 P03B 075 29,  182/N-QGG/2008/0 and
the 2007-2010 research project N202 n081 32/1844, the Foundation
for Polish Science grant ``Master'', The George A. and Margaret M.
Downsbrough Endowment and the Eberly research funds of Penn State.

\end{document}